\documentclass[twocolumn,prl,floatfix,superscriptaddress,showpacs]{revtex4-1}

\usepackage{amsmath}
\usepackage{amsfonts}
\usepackage{amsbsy}
\usepackage{xcolor}
\usepackage{soul}
\usepackage{graphicx}
\usepackage{epsfig}
\usepackage{epstopdf}
\usepackage{hyperref}

\newcommand{\pol}{\hat{\bf e}}
\newcommand{\rv}{{\bf r}}
\newcommand{\Ev}{{\bf E}}
\newcommand{\Dv}{{\bf D}}
\newcommand{\Pv}{{\bf P}}

\newcommand{\dv}{{\bf d}}

\newcommand{\eo}{\epsilon_0}
\newcommand{\beq}{\begin{equation}}
\newcommand{\eeq}{\end{equation}}
\newcommand{\bea}{\begin{eqnarray}}
\newcommand{\eea}{\end{eqnarray}}
\newcommand{\BEQAL}{\begin{align}}
\newcommand{\EEQAL}{\end{align}}
\newcommand{\EQREF}[1]{Eq.~(\ref{#1})}
\newcommand{\comment}[1]{{}}

\newcommand{\<}{\langle}
\renewcommand{\>}{\rangle}
\renewcommand{\(}{\left(}
\renewcommand{\)}{\right)}

\newcommand{\brho}{{\boldmath\hbox{$\varrho$}}}
\newcommand{\commentout}[1]{{}}

\begin{document}
\draft
\preprint{}
\title{Emergence of correlated optics in one-dimensional waveguides for classical and quantum atomic gases}
\author{Janne Ruostekoski}
\affiliation{Mathematical Sciences, University of Southampton, Southampton SO17 1BJ, United Kingdom}
\author{Juha Javanainen}
\affiliation{Department of Physics, University of Connecticut, Storrs, Connecticut 06269-3046}
\date{\today}
\begin{abstract}
We analyze the emergence of correlated optical phenomena in  the transmission of light through a waveguide that confines classical or ultracold quantum degenerate atomic ensembles.
The conditions of the correlated collective response are identified in terms of atom density, thermal broadening, and photon losses by using stochastic Monte-Carlo simulations and transfer matrix methods of transport theory.
We also calculate the ``cooperative Lamb shift'' for the waveguide transmission resonance, and discuss line shifts that are specific to effectively one-dimensional waveguide systems.
\end{abstract}
\maketitle

Confining the light in a region comparable with the atomic scattering cross section can considerably enhance atom-light coupling and lead to new regimes of light-matter interactions. Guided modes of 1D waveguides~\cite{kimblenaturecom} and nanofibers~\cite{Mitsch,Rauschenbeutel} open up new avenues of optical physics where light propagation could potentially be employed in high-precision spectroscopy~\cite{lambdickespec}, quantum networks, light circuitry, and quantum switches~\cite{lukintransistor,Volz14,Fan1}. For instance, superradiance of atoms confined inside a photonic crystal waveguide was recently reported~\cite{kimblesuper}, and 1D waveguides support long-range light-mediated interactions with also the possibility of creating novel quantum many-body phases~\cite{cirackimble} for atoms and light. Atomic waveguides also have close analogies in other
1D electrodynamics realizations, such as with different nanoemitter systems~\cite{Petersen,Lodahl_qd,capasso_cherenkov,Sapienza}, surface plasmon nanowires~\cite{lukintransistor}, coupled-cavity QED~\cite{Hartmann}, and superconducting transmission lines~\cite{Wallraff,Fan1}.

In anticipation of the importance of many-atom physics in waveguide systems,
we raise here the question: when do the atoms respond to light independently, as in an ordinary optical medium, and when is the response correlated?
In an ideal 1D waveguide the light emitted by an atom travels unattenuated with a constant amplitude, and one might think that the corresponding infinite-range radiative dipole-dipole (DD) interaction sets up global correlations between the atoms. Maybe surprisingly it is not so, and the (line) density of the atoms makes a difference. We find that for randomly distributed atoms the point of demarkation is the wave number of resonant light $k$. At low density, the propagation delays in the multiple scattering of light between the atoms are sufficiently random that the atoms, in fact, transmit light basically independently, whereas at high density the propagation delays, being small, cannot be altogether random, and light-induced correlations emerge. This is an interesting analogy with 3D systems, where it has been found that when the typical interatomic separation is comparable or less than $1/k$, the atomic gas can exhibit a correlated response and the traditional electrodynamics fails~\cite{Javanainen2014a}. An unambiguous observation of correlated optics has proven elusive in 3D gases, however, so 1D systems may offer a promising alternative.

The onset of emergent correlations is characterized not only in terms of atom density, but also imperfections of the waveguide such as the fraction of light radiated by the atoms that leaks out of the waveguide, and Doppler broadening of the resonance resulting from the thermal velocity distribution.
We point out that the collective behavior in 1D entails a shift of the resonance line proportional to the line density of atoms. The 3D analog here are line shifts proportional to volume density, which are a well-known complication in high-precision spectroscopy.
For quantum degenerate (not randomly distributed) atoms the light-induced correlations remain stronger and we find, e.g., that fermionic
atom statistics leads to resonance linewidth narrowing and suppressed superradiance.

In the following, for our analysis we develop a quantum-mechanical theoretical framework for light propagation in classical and quantum degenerate atomic ensembles in 1D waveguides.
Classical electrodynamics simulations provide exact solutions within the model of two-level, weakly excited, stationary atoms.
An especially elegant representation of light-induced
correlations is obtained using transfer matrices where we adapt theoretical methods of localization analysis in transport phenomena~\cite{anderson1d} that were originally developed for 1D electric conductivity.

We assume a narrow waveguide where the forward and backward propagating modes are determined by the wavenumber $q$ and the polarization components $\hat{\bf u}_{q\sigma}(\brho)$,
with $\int d^2\varrho\, \hat{\bf u}^*_{q\sigma}(\brho)\cdot\hat{\bf u}_{q\sigma}(\brho)=1$; the propagation direction is denoted by $x$, and the transverse coordinate by $\brho$.
We describe the interactions of light and atoms in the \emph{length} gauge that
is obtained by the Power-Zienau-Woolley transformation~\cite{PowerBook}.
The positive frequency component of the electric displacement $\Dv^+(\rv)$ then reads
\beq
\Dv^+(\rv)= \sum_{q,\sigma} \zeta_q \hat{\bf u}_{q\sigma}(\brho)\, \hat a_{q\sigma} e^{i q x},\quad
\zeta_q=\sqrt{\frac{\hbar \epsilon_0\omega_q}{2 L}}\,,
\label{eq:Dquantisation}
\eeq
where the mode frequency, the photon annihilation operator, and the quantization length are denoted by $\omega_q$, $\hat a_q$, and $L$, respectively. Due to the spatial confinement~\cite{Petersen,Mitsch}, the summation over the  polarizations $\sigma$ generally involves both transverse and longitudinal components.
The electric field $\Ev^+$ in the waveguide may then be integrated using standard techniques of quantum optics~\cite{milonniknight}, and
expressed as a sum of the incident field $\Dv^+_F$ and the scattered field,
\begin{align}
\eo \Ev^+({\bf r})& = \Dv^+_F({\bf r}) +
\int d^3r'\,
{\sf G}(\rv,\rv')\,\Pv^+({\bf r}')\,,
\label{eq:MonoD}\\
{\sf G}(\rv,\rv') &= {ik\over2}\,e^{ik|x-x'|}  {\sf M}(k;\brho,\brho') \,,
\label{eq:G}
\end{align}
where the electric polarization $\Pv^+(\rv) =\sum_{ge}  \Pv_{ge}^+(\rv)= \sum_{ge}  \dv_{ge} \,\psi^\dagger_{g}(\rv)\psi_{ e}(\rv)$ acts as a radiation source.
We have introduced the atomic field operators for the electronic ground and excited states $\psi_{g}(\rv)$ and  $\psi_{ e}(\rv)$, with the Zeeman levels included in the indices $g$ and $e$  when applicable, and the dipole matrix element $\dv_{ge}={\cal D}\sum_\sigma \pol_\sigma {\cal C}_{g,e}^{(\sigma)}$ for the atomic transition $|e\> \rightarrow |g\>$. Here the summation is over the circularly polarized unit vectors $\pol_\sigma$, ${\cal C}_{g,e}^{(\sigma)}$ denote the Clebsch-Gordan coefficients, and ${\cal D}$ is the reduced dipole matrix element.
The polarization of the scattered light is determined by the tensor $ {\sf M}(k;\brho,\brho')$ that accounts for the projection to the transverse mode $\hat{\bf u}_{q\sigma}(\brho)$ and the radial position $\brho'$ of the radiating atom.
For instance, if the atoms with a complex level structure are trapped outside a nanofiber where the gradient of the evanescent field in the radial direction is large, the contribution of the longitudinal polarization can be significant leading to `chiral', axial-direction-dependent emission~\cite{Petersen,Mitsch}.
We have assumed that there is a dominant frequency $\Omega=kc$ of the driving light and, for simplicity of notation, here and in the rest of the paper we have written all operators
in the  ``slowly varying'' picture by explicitly factoring
out the dominant frequency component; $\Dv^+\rightarrow e^{-i\Omega t} \Dv^+$, $ \Pv^+\rightarrow e^{-i\Omega t} \Pv^+$, etc.
Owing to a single-mode nature of the waveguide, the radiation kernel ${\sf G}(\rv,\rv')$ has the form of a 1D propagator~\cite{Javanainen1999a} that does not lead to attenuation of the light propagating in the axial direction.

In order to solve the scattered field in Eq.~\eqref{eq:MonoD}, the equation of motion for $\Pv^+_{ge}$ can be derived analogously to the full field-theoretical
treatment of the 3D electrodynamics~\cite{Ruostekoski1997a}, even while keeping the general hyperfine-level and polarization structure. However, in the following  we assume that the atoms are tightly confined in the radial direction at the center of the waveguide ($\varrho=0$),
such that the effect of the radial dependence of the field mode on the atoms may be ignored, and that  it is sufficient to consider scalar equations for each polarization component.
We may then consider a two-level system,
and replace $|\hat{\bf u}_q(\varrho\simeq0)|$ by the inverse of the characteristic length scale of the radial light mode confinement. We take the radial light intensity profile to be a Gaussian with the $1/e$ width $\xi_\varrho$, such that $u(\varrho\simeq0)=1/\sqrt{\pi}\xi_\varrho$. Furthermore, we integrate over the radial dependence of the atomic polarization and density and, for simplicity of notation, assume that they have the same radial profile.

This results in an effective 1D theory (Appendix) with the replacement $\pi\xi_\varrho^2 {\Dv}^+_F(\rv)\rightarrow \tilde{D}^+_F(x)$, etc., where the radiation kernel ${\sf G}$ becomes a Green's function for the 1D Helmholtz equation $G(x-x')= ik e^{ik|x-x'|}/2\pi\xi_\varrho^2$~\cite{BOR99}.
The scattered field then depends on the scalar polarization $P^+={\cal D} {\cal C}_{ge}\psi^\dagger_g \psi_e$, and in the limit of low light intensity we obtain for the expectation value of the steady-state polarization $P_1\equiv \<P^+\>$
\beq
{P}_1(x) = \alpha\rho \tilde{D}^+_F(x)+\eta_\delta \int
dx' e^{ik|x-x'|} {P}_{2}(x;x')\,,
\label{eq:p1}
\eeq
where we have defined a single-atom polarizability in a 1D waveguide as $\alpha=-2\gamma_w/[ k(\delta+i\gamma_t)]$ in terms of the radiative linewidth $\gamma_t=\gamma_l+\gamma_w$ that depends on the radiative losses out of the waveguide $\gamma_l$ and on the decay rate into the waveguide $
\gamma_w = {k{\cal D}^2 / 2\pi \xi_\varrho^2\hbar\eo}$.
The detuning of
$\Omega$ from the atomic
resonance is denoted by $\delta$,
the atom density by $\rho$, and $\eta_\delta\equiv
{\gamma_w/(i\delta-\gamma_t)}=i\alpha k/2$.

Now, the polarization $P_1$ depends on the two-atom correlation function $P_2(x;x')=\< \psi_g^\dagger(x) P^+(x') \psi_g(x)\>$.
Analogously, for $P_2$ we obtain the steady-state solution
\begin{align}
{P}_2(x_1;x_2) &= \alpha\rho(x_1,x_2) \tilde{D}^+_F(x_2) +\eta_\delta
e^{ik|x_1-x_2|} {P}_{2}(x_2;x_1)\nonumber\\
&+\eta_\delta \int
dx_3\, e^{ik|x_2-x_3|} {P}_{3}(x_1,x_2;x_3)\,,
\label{eq:p2}
\end{align}
where $\rho(x_1,x_2)$ denotes the ground-state atom pair correlation function that in the limit of low light intensity is unaffected by the driving light.
In \EQREF{eq:p2}, $P_2$ depends on the three-body correlation function $P_3(x_1,x_2;x_3)$
for polarization at $x_3$, and ground-state atom densities at $x_1$ and $x_2$.
The three-atom correlation function $P_3$ in turn depends on four-atom correlations, etc., leading to the hierarchy of equations for the correlation functions.

After the rescaling of the electromagnetic fields, the cooperative response of atoms in a 1D waveguide
closely resembles a 1D model electrodynamics~\cite{Javanainen1999a} that described a hypothetical system consisting of continuously distributed 2D planes of atomic dipole moments in which the radiators are discrete only in the direction of the light propagation. The one change is that we have specifically introduced the loss rates due to spontaneous emission out of the waveguide, and thus address an actual experimentally relevant physical system.

The second term in Eq.~\eqref{eq:p2} describes repeated exchanges of a photon between atoms at $x_1$ and $x_2$. Such \emph{recurrent} scattering processes~\cite{Morice1995a,Ruostekoski1997a} between nearby atoms are responsible for light-induced correlations between the atoms.
The hierarchy can be solved exactly by means of stochastic simulations where the positions of the atoms are sampled from a probabilistic ensemble that corresponds to the position correlations between the atoms in the absence of the driving light~\cite{Javanainen1999a}. In each stochastic realization of discrete atomic positions $\{x_1,x_2,\ldots,x_N\}$, we solve for the coupled set of classical electrodynamics equations for point dipoles that account for the polarization density $\sum_j \mathfrak{ P}^{(j)}\delta(x-x_j)$, where $\mathfrak{ P}^{(j)}$ denotes the excitation dipole of the atom $j$. The coupled-dipole equations for the steady-state solution in the present system read
\beq
\mathfrak{ P}^{(j)} = \alpha \tilde D^+_F(x_j) + \eta_\delta \sum_{l\neq j} e^{ik|x_j-x_l|}\mathfrak{ P}^{(l)}\,,
\label{eq:classicaled}
\eeq
where each dipole amplitude is driven by the incident field and the scattered field from all the other $N-1$ dipoles. Once all $\mathfrak{ P}^{(j)}$ are calculated, the scattered fields in each realization may be obtained from $\eo E^+_{\rm sc}(x)=ik\sum_l \exp (ik|x-x_l|) \mathfrak{ P}^{(l)}/2$, and the total field equals the incoming field plus the scattered field. Finally, evaluating the ensemble average over many sets of atomic positions with the correct probability distribution generates the exact solution to the optical response of stationary atoms with a given atom statistics in the low-excitation limit~\cite{Javanainen1999a}.
The applicability of the simulations extends beyond atoms, since similar methods can be employed, e.g., in
nanoresonator systems~\cite{JenkinsLongPRB,CAIT}.

The coupled dynamics for the atoms and light confined inside the waveguide exhibits characteristic behavior of 1D electrodynamics; for instance, for an exact resonant excitation the first atom can reflect all the light~\cite{Javanainen1999a}. The single-atom transmission and reflection amplitudes equal
\beq
t^{(1)} = {(\gamma_w-\gamma_t)+i\delta\over i\delta-\gamma_t},\quad
r^{(1)} ={\gamma_w\over i\delta-\gamma_t}\,,
\label{eq:singletrans}
\eeq
with the single-atom power transmission and reflection coefficients given by $T^{(1)} =|t^{(1)} |^2$ and $R^{(1)}= |r^{(1)} |^2$.
The total reflection occurs for suppressed photon losses, $\gamma_w=\gamma_t$. It is a generic phenomenon of 1D  scattering -- e.g., the origin of the Tonks gas behavior of impenetrable bosonic atoms in strongly confining 1D traps~\cite{Olshanii}.

The hierarchy of equations represents light-induced correlations between the atoms that result from the recurrent scattering. In mean-field theory (MFT) such correlations are ignored. Next, we construct MFT solutions by specifically neglecting all recurrent scattering events, indicating that no photon scatters more than once by the same atom. For $N$ atoms this means that each atom in a row simply passes on the same fraction of light and the transmission coefficient is given  in terms of the single-atom transmission amplitude (\ref{eq:singletrans}) as
$ t^{(N)}_{\rm mft}  = \< t^{(1)}\>^N$.
Here we show that the MFT result can dramatically fail when light-induced correlations between the atoms become important.
To model an atom cloud we solve the light transmission through the waveguide when the atomic positions are stochastically distributed. A cold classical atomic ensemble or an ideal Bose-Einstein condensate can be analyzed by sampling independent random atomic positions,
while simulations
in the quantum degenerate regime require one to synthesize a stochastic ensemble of atomic positions that generates the proper position correlations. We illustrate the latter by considering metrologically important fermionic correlations (a zero-temperature fermionic gas or an impenetrable bosonic Tonks gas) in which case the stochastic ensemble in each run is generated with the Metropolis algorithm~\cite{Ceperley}.
\begin{figure}
\includegraphics[width=0.95\columnwidth]{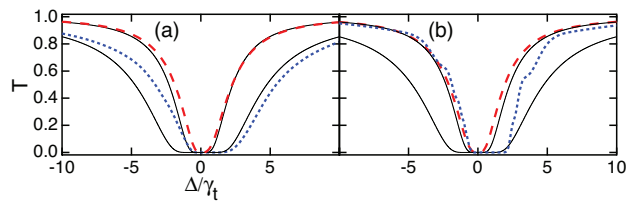}
\vspace{-0.4cm}
\caption{
Light-induced correlation effects between the atoms as evinced by the power transmission coefficient through the waveguide $T$.
The full numerical solution at atom densities $\rho k^{-1}=2$ (dashed line) and 8 (dotted line) vs the corresponding MFT results (solid lines) for $T$ as a function of the detuning of the incident light from the single-atom resonance. The increased deviations  indicate the growing importance of recurrent scattering. (a) Classical atoms that are uncorrelated before the light enters the sample; (b) fermionic atoms.   }
\label{fig:mftvsfull}
\end{figure}

In Fig.~\ref{fig:mftvsfull} we show the exact simulation results with the corresponding MFT solution for different atom densities.
At low densities the exact solution -- that by definition fully incorporates all light scattering -- coincides with the  approximate MFT analysis that neglects all recurrent scattering processes and treats the atoms as independent. As the atom density gets higher the MFT solution becomes increasingly inaccurate, indicating the emergence of light-induced correlations between the atoms. It is perhaps surprising that an ensemble with a uniform density--such as a random distribution of classical atoms or, alternatively, a delocalized condensate wavefunction--exhibits correlated optics. Such a system mimics  a continuous optical medium with a uniform refractive index~\cite{BOR99}, yet the light is still able to establish correlations between atoms, violating the standard continuous-medium optics.

The presence or absence of correlations due to recurrent scattering processes, as in Fig.~\ref{fig:mftvsfull}, may be understood by considering fluctuations in the light propagation phases between the adjacent atoms. We adapt localization analysis using transfer matrices~\cite{anderson1d}, originally introduced for 1D electric conductivity (Appendix). The MFT description $t^{(N)}_{\rm mft}$ becomes accurate whenever the transmission amplitude for $N$ atoms at random positions factorizes into independent-atom contributions, $\<t^{(N)}_{1,2,\ldots,N}\> =\< t^{(1)} \>^N$, where $t^{(1)}$ is given by \EQREF{eq:singletrans}. (Alternatively, we can describe
the factorization in terms of the $N$-atom optical thickness (Appendix).)
To determine the validity of MFT it is sufficient to consider a two-atom subsystem that can be recursively generalized to the $N$-atom case. For two atoms we find
\beq
\< t_{12}^{(2)}\> = \big\<  {  t_2^{(1)} t_1^{(1)}\over 1 - \sqrt{R_1^{(1)} R_2^{(1)}} e^{i\phi}} \big\>\,.
\label{eq:average1a}
\eeq
The denominator can be represented as a geometric series where each subsequent term includes one additional recurrent scattering event between the atom pair (Appendix).
The phase $\phi=\varphi_1+\varphi_2+2k x_{12}$ ($x_{12}=x_2-x_1$) consists of the light propagation phase $2k x_{12}$ from atom 1 to atom 2 and back, and the contributions $\varphi_j=\arctan(\delta_j/\gamma_t)$  from the atomic reflectance that are sensitive to the detunings of the driving light from the atomic resonance $\delta_j$.

As may be seen by doing the average on the right-hand side of Eq.~(\ref{eq:average1a}), MFT results, with
the decoupling of the transmission amplitudes between the two atoms $\< t_{12}^{(2)}\> \simeq  t_{1}^{(1)} t_{2}^{(1)} $,
if the propagation phases are distributed evenly over $[0,2\pi)$ (Appendix). In fact, when the density is sufficiently low, $\rho \ll \pi/k$, so that for the characteristic interatomic separation $\ell= 1/\rho$ we have the propagation phase $2\ell k \gg 2\pi$, for random atomic positions the propagation phase for two adjacent atoms is distributed approximately evenly over $[0,2\pi)$, and light-induced correlations are suppressed. At higher $\rho$ the interatomic separation between the adjacent atoms is no longer large enough for the propagation phases to be random. Consequently, the light-induced correlations are not canceled out and we observe deviations from MFT. Analogously to the 3D case~\cite{Javanainen2014a} the relevant length scale is $1/k$.

The cancelation of the effects of recurrent scattering may also  occur at high densities in an inhomogeneously-broadened hot atom vapor due to the Doppler shifts of the resonance frequencies.
In Eq.~(\ref{eq:average1a}) the detunings $\delta_j$ also appear in the single-atom transmission $t_j^{(1)}$ and reflectance $r_j^{(1)}$ [\EQREF{eq:singletrans}].
If $\delta_j$  have a sufficiently broad distribution, then averaging over the velocity distribution of the atoms eliminates the recurrent scattering, and MFT becomes valid.
The relevant energy scale for the Doppler broadening is the resonance linewidth of the atoms, and MFT is valid
whenever the temperature is high enough such that $k\sqrt{k_BT/m}\gg \gamma_t$ (Appendix).

In anticipation of ultracold many-body physics in hybrid waveguide systems, we extend calculations to quantum degenerate ensembles where the atoms
still form an optical medium with a uniform density but they are no longer randomly distributed.
If the atomic positions are correlated as a result of fermionic fluctuations,  the short-range Fermi repulsion between the atoms creates an additional bias in the distribution of the propagation phases. We then find more dramatic violation of the MFT predictions [Fig.~\ref{fig:mftvsfull}].
 Evidently the phase in the denominator of Eq.~\eqref{eq:average1a} is less easily randomized away, leading to the strengthening of correlations in light propagation, resonance line narrowing, and suppressed superradiance.

\begin{figure}
\includegraphics[width=0.99\columnwidth]{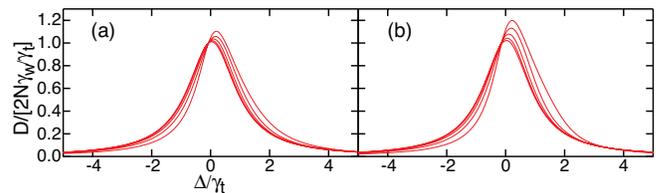}
\vspace{-0.8cm}
\caption{Scaled optical thickness $- \<\ln T_{12\ldots N}^{(N)} \>/(2N\gamma_w/\gamma_t)$ as a function of the detuning of light from the
atomic resonance for  $\gamma_w/\gamma_t=0.4$, $0.2$, $0.1$, $0.05$, and $0.025$ (curves from top to bottom) for (a) initially uncorrelated classical atoms; (b) degenerate fermionic atoms ($L=2\lambda$, $N=32$).
For $\gamma_w/\gamma_t\rightarrow 0$, all cases converge to the MFT result of $\gamma_t^2/(\gamma_t^2+\delta^2)$. }
\label{SEQUENCES}
\end{figure}

Even though photonic crystal waveguide experiments strive toward low loss rate of photons from the waveguide, nanofiber systems typically have
$\gamma_w/\gamma_t\ll 1$. With high losses, the number of multiple scattering events any single photon can undergo inside the waveguide is limited.
To leading order in $\gamma_w/\gamma_t$, we obtain for the two-atom transmission $
T_{12}^{(2)} \simeq  T_1^{(1)} T_2^{(1)} + {\cal O} (\gamma_w^2/\gamma_t^2) $, indicating the recovery of the MFT results. In Fig.~\ref{SEQUENCES} we show the numerically simulated light transmission for different loss rates. In the limit $\gamma_w/\gamma_t\rightarrow 0$ the curves converge toward the MFT result, but notable deviations can be identified even in the case of strong losses.

\begin{figure}
\includegraphics[width=0.99\columnwidth]{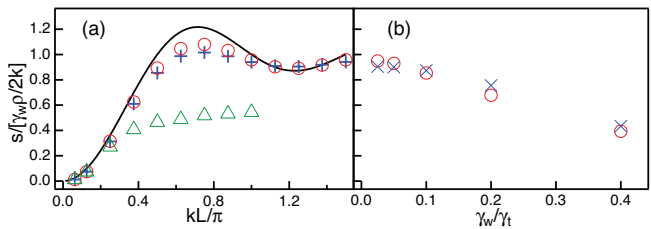}
\vspace{-0.6cm}
\caption{Frequency shift  $s$ of the maximum of the  light intensity transmitted through the waveguide. (a) The shift (solid line) as predicted by the ``cooperative Lamb shift'' model~(\ref{COLLAMB}), and the full numerics (at $\rho=32\, k/\pi$) for $\gamma_w/\gamma_t=0.01$ (circles), 0.02 (crosses), and 0.1 (triangles) as a function of sample thickness. (b) Variation of line shift with waveguide loss rate. These are the shifts of the curve maxima in Fig.~\ref{SEQUENCES} for classical (circles) and fermionic (crosses) atoms.}
\label{fig:shift}
\end{figure}

We calculated the MFT results by treating the transmission of each atom independently and then taking the product of the transmissions of the independent atoms.
We may also neglect the recurrent scattering and light-induced correlations between the atoms directly in the hierarchy of equations by  factorizing the correlation function ${P}_{2}(x;x')\simeq \rho(x) {P}_1(x')$ in \EQREF{eq:p1}. This truncates the hierarchy, provides closed equations from which ${P}_1$ and the scattered fields may be solved, and leads to an effective-medium MFT.
It was shown~\cite{Javanainen2014a} in a 3D system that in the low atom-density limit the factorization reproduces the ``cooperative Lamb shift'' (CLS) that
Friedberg {\it et al}.~\cite{Friedberg1973} have calculated for various 3D geometries of atomic ensembles. The experimental measurement of CLS has attracted considerable interest with nuclei~\cite{ROH10}, ions \cite{Meir13}, and CLS was recently qualitatively verified in hot~\cite{Keaveney2012} and low-density~\cite{Roof16} atomic vapors. In systems where the light-induced correlations between the emitters become strong, CLS prediction can fail~\cite{Javanainen2014a}. Here the factorization ${P}_{2}(x;x')\simeq \rho(x) {P}_1(x')$ in \EQREF{eq:p1} gives  CLS of the 1D waveguide in the limit of asymptotically small density (Appendix)
\beq
\Delta_{\rm CLS} ={\gamma_w\rho\over 2k}\left( 1-\frac{\sin 2Lk}{2Lk}\right) .
\label{COLLAMB}
\eeq
The oscillatory behavior corresponds to the etalon effect due to the sample thickness.
In 1D physical and practical constraints conspire to make it difficult to verify the result~(\ref{COLLAMB}) numerically, but for judiciously chosen parameters we get close [Fig.~\ref{fig:shift}(a)]. Moreover, in numerical computations the  resonance shift is found to be on the order of  ${\gamma_w\rho/ (2k)}$ for a wide range of parameters. This is illustrated in Fig.~\ref{fig:shift}(b) for the line shifts in Fig.~\ref{SEQUENCES}.

In conclusion, nanophotonic waveguides can naturally enhance light-mediated collective response in atomic ensembles. 
Collective optical phenomena find applications, e.g., in engineering superradiance~\cite{kimblesuper}, narrow spectral linewidths~\cite{CAIT,fanoreview}, enhanced extinction~\cite{CAIT,Bettles_prl16}, subwavelength excitations~\cite{LemoultPRL10}, lasers~\cite{Bohnet2012a}, controlling line shifts~\cite{Javanainen2014a,Keaveney2012,ROH10,Meir13,Jenkins_thermshift}, and in 1D waveguides in the studies of Anderson-localized modes of light~\cite{Sapienza}.
Here we analyzed light transmission through an atomic ensemble in a waveguide. The light-induced correlations due to recurrent scattering were identified both in simulations and in a transfer matrix analysis. The validity of MFTs was characterized in terms of atom density, thermal broadening, and photon loss rate. We also pointed out that quasi-1D waveguide systems may exhibit perhaps unexpected frequency shifts -- an observation that may be relevant in sensing and metrology~\cite{lambdickespec}.

\begin{acknowledgments}
We acknowledge support from NSF, Grant Nos. PHY-0967644 and PHY-1401151, and EPSRC.
\end{acknowledgments}

\appendix

\setcounter{equation}{0}
\setcounter{figure}{0}
\renewcommand{\theequation}{A\arabic{equation}}
\renewcommand{\thefigure}{A\arabic{figure}}

\section{Appendices}

\subsection{One-dimensional electrodynamics}

\subsubsection{Hierarchy of equations of motion}

For instance, in situations where the atoms are strongly confined close to the centre of the waveguide or when an effective two-level system is obtained from the $J=0\rightarrow J'=1$ system, we may consider the 1D scalar electrodynamics for the coupled system of atoms and light by renormalizing the fields, as $ \pi\xi_\varrho^2 \Ev^+(\rv)\rightarrow \tilde E^+(x)$, $ \pi\xi_\varrho^2 \Dv^+_F(\rv)\rightarrow \tilde D^+_F(x)$. The total electric field amplitude is the sum of the incident and the scattered fields
\beq
\eo \tilde E^+(x) = \tilde D^+_F(x) + {ik\over2} \int dx'\, e^{ik|x-x'|}\,P^+(x')\,,
\label{eq:MonoD1d}
\eeq
where $k=\Omega/c$ and $\Omega$ is the frequency of the incident driving field.
For the coherently scattered light we take the expectation values  of Eq.~(\ref{eq:MonoD1d}) and assume that the monochromatic incident field is in a coherent state. Noting that $ie^{ik|x|}/(2k)$ is the the Green's
function of the 1D Helmholtz differential operator,
\begin{equation}
(\nabla^2+k^2)\langle{\tilde D}^+_F\rangle = 0, \quad
(\nabla^2+k^2)\, {i\over2k}\,e^{ik|x|} = -\delta(x)\,,
\end{equation}
we can transform the integral equation (\ref{eq:MonoD1d}) to a differential equation
\begin{equation}
(\nabla^2+k^2)\langle{\tilde D}^+\rangle =
\nabla^2\langle{ P}^+\rangle\,,
\label{eq:FEX}
\end{equation}
which is the 1D Maxwell's wave equation in a polarizable medium~\cite{BOR99}.

The atomic polarization acts as a radiation source in Eq.~(\ref{eq:MonoD1d}) and in second quantization is expressed as $P^+(x) = {\cal C}_{ge} {\cal D}\,\psi^\dagger_{g}(x)\psi_{ e}(x)$ in terms of
the atomic field operators for the electronic ground $\psi_{g}(x)$ and excited $\psi_{ e}(x)$ states, the Clebsch-Gordan coefficient ${\cal C}_{ge}$, and the reduced dipole matrix element ${\cal D}$.
In order to solve the coupled theory for atoms and light, we need the Heisenberg's equations of motion for the atomic polarization operator $P^+(x)$. This can be derived from the interaction term
between the atomic polarization and the electric displacement field $P(x) \tilde D(x)/\eo$.
For the expectation value ${P}_1(x)=\<{P}^+(x)\>$ we obtain 
\begin{align}
\dot {P}_1(x) =& (i\delta -\gamma_t) {P}_1(x)+ {2i \gamma_w\over k}\rho \tilde{D}^+_F(x)\nonumber\\
&-\gamma_w \int
dx' e^{ik|x-x'|} {P}_{2}(x;x')\,.
\label{p1app}
\end{align}
Here $\delta=\Omega-\omega_0$ denotes the detuning of the light frequency $\Omega$ from atomic resonance frequency $\omega_0$ and the radiative inewidth is given by
\beq
\gamma_t=\gamma_l+\gamma_w\,,
\eeq
 where $\gamma_l$ is the radiative loss rate from the waveguide and
 \beq
 \gamma_w= k{\cal D}^2  / 2\pi \xi^2_\rho \hbar\eo
 \eeq
the decay rate into the waveguide, with the characteristic radial length scale $\xi_\rho$.
Equation~\eqref{p1app} is derived in the low light intensity limit, i.e., to first order of the amplitude $D$, and correspondingly keeping terms that include at most one of the fields $D,\psi_e,\psi_e^\dagger$~\cite{Ruostekoski1997a}. In the low light intensity limit the ground-state atom density $\rho$ is not changed by the driving light, and has the same value as before the light enters the sample.
Light-mediated dipole-dipole interactions induce correlations between the atoms that depend on their positions. Polarization depends on the two-atom correlation function
\beq
P_2(x_1;x_2)\equiv \<\psi^\dagger_g(x_1)
 P^+(x_2) \psi_g(x_1) \>\,.
\eeq
$P_2$ describes correlations between a ground-state atom at $x_1$ and the polarization at $x_2$.
Similarly, the equation for $P_2$ is coupled to a three-atom correlation function
\begin{align}
\dot {P}_2(x_1;x_2) =& (i\delta -\gamma_t) {P}_2(x_1;x_2) -\gamma_w
e^{ik|x_1-x_2|} {P}_{2}(x_2;x_1)\nonumber\\
&+ {2i \gamma_w\over k}\rho(x_1,x_2) \tilde{D}^+_F(x_2)\nonumber\\
&-\gamma_w \int
dx_3\, e^{ik|x_2-x_3|} {P}_{3}(x_1,x_2;x_3)\,,
\end{align}
where
\beq
P_3(x_1,x_2;x_3)\equiv \<\psi^\dagger_g(x_1)
\psi^\dagger_g(x_2) P^+(x_3) \psi_g(x_2) \psi_g(x_1) \>\,,
\eeq
represents a polarization density at $x_3$, given that there are ground-state atoms at $x_1$ and $x_2$.
The three-atom correlation function $P_3$ in turn depends on four-atom correlations, which are coupled to
five-atom correlations, etc., leading to the hierarchy of equations for the correlation functions that involve the polarization and an increasing number of ground-state atom densities.

The ground-state atom pair correlation function
\beq
\rho(x_1,x_2)\equiv \<\psi^\dagger_g(x_1)\psi^\dagger_g(x_2)\psi_g(x_2) \psi_g(x_1) \>\,,
\eeq
in the low light-intensity limit is unaffected by the driving light and corresponds to the initial correlations between the atoms in the absence of the driving light.

Generally, we define normally ordered mixed polarization-atom density
correlation functions and correlation functions of density for ground
state atoms as
\begin{align}
&{P}_l(x_1,\cdots,x_{l-1};x_l) \nonumber\\&\equiv
\langle \psi^\dagger_g(x_1)\cdots\psi^\dagger_g(x_{l-1})
{P}^+(x_l) \psi_g(x_{l-1})\cdots
\psi_g(x_1)\rangle,\label{PLQ}\\&\rho_l(x_1,\cdots,x_l) \equiv \langle
\psi_g^\dagger(x_1)\cdots
\psi_g^\dagger(x_l) \psi_g (x_l)\cdots \psi_g (x_1)
\rangle\,,
\label{RRR}
\end{align}
with $l=1,2,\ldots$.  The hierarchy of equations for the correlation
functions may then be written in the following form
\begin{eqnarray}
\lefteqn{\dot{P}_l(x_1,\cdots,x_{l-1};x_l)=
(i\delta-\gamma_t) {P}_l(x_1,\cdots,x_{l-1};x_l)}\nonumber\\
&&-\gamma_w\sum_{k=1}^{l-1}e^{ik|x_l-x_k|} {
P}_l(x_1,\cdots,x_{k-1},x_{k+1},\cdots,x_l;x_k) \nonumber\\
&&+\,{2i\gamma_w\over k}\,
\rho_l(x_1,\cdots,x_l)\,{\tilde D}^+_F(x_l)\nonumber\\ &&-\gamma_w\int
dx_{l+1}\,e^{ik|x_l-x_{l+1}|} {P}_{l+1}(x_1,\cdots,x_l;x_{l+1})\,.
\label{eq:hie}
\eea
The terms in the sum on the right-hand side of Eqs.~{(\ref{eq:hie})}
represent recurrent scattering processes in which the $l$ atoms at $x_1,\ldots,x_l$ repeatedly
exchange photons. Such processes are the microsocopic mechanism for
collective linewidths and line shifts. The integral stands for a
process in which yet another atom shines its light on the atom at $x_l$.

\subsubsection{Stochastic simulations}

In the limit of low light intensity, the hierarchy of equations of motion for the correlation functions \eqref{eq:hie} represents coupled weakly excited linear atomic dipoles.
Although the recurrent scattering can induce strong correlations between the atoms, in a two-level atomic ensemble the hierarchy can, nevertheless, be entirely understood by classical electrodynamics. Indeed, the hierarchy of equations for the correlation functions can be solved exactly by means of classical stochastic simulations where the positions of the atoms are sampled from a probabilistic ensemble that corresponds to the position correlations between the atoms in the absence of the driving light. The technique fully accounts for the recurrent scattering processes to all orders.

If the atoms are initially uncorrelated before the light enters the medium, as for classical atoms or for an ideal Bose-Einstein condensate, the stochastic position of each atom is sampled independently from the positions of the other atoms. For the zero-temperature fermionic position correlations (for an ideal gas of fermionic atoms or for a strongly interacting Tonks gas of bosonic atoms) the initial positions of the atoms are given by the joint probability distribution equal to the absolute square of the  Fermi-Dirac many-atom wave function. The correlated positions may then be sampled using the Metropolis algorithm~\cite{Ceperley}.

In each stochastic realization the optical response of a system of
classical electrodynamics equations for point dipoles is solved exactly for a fixed set of atomic positions.
The dynamics of the hierarchy of correlation functions \eqref{eq:hie}
can then be reproduced to all orders by  ensemble-averaging over many realizations of spatial positions that are stochastically sampled from the joint many-body probability
distribution~\cite{Javanainen1999a}. 

Specifically, in each stochastic realization of discrete atomic positions $\{x_1,x_2,\ldots,x_N\}$, we solve for the coupled set of point dipoles that account for the polarization density $\sum_j \mathfrak{ P}^{(j)}\delta(x-x_j)$, where $\mathfrak{ P}^{(j)}$ denotes the excitation dipole of the atom $j$. The coupled-dipole equations for the steady-state solution in the present system read
\beq
\mathfrak{ P}^{(j)} = \alpha \tilde D^+_F(x_j) + \eta_\delta \sum_{l\neq j} e^{ik|x_j-x_l|}\mathfrak{ P}^{(l)}\,,
\label{eq:classicaledb}
\eeq
where each dipole amplitude is driven by the incident field and the scattered field from all the other $N-1$ dipoles. Once all $\mathfrak{ P}^{(j)}$ are calculated, we may obtain in each realization the expression for the scattered light from the induced dipole excitations
\beq
\eo \tilde E^+(x)= \tilde D^+_F(x)+ {ik\over2} \sum_l e^{ik|x-x_l|} \mathfrak{ P}^{(l)}\,.
\label{eq:pointlight}
\eeq
Finally, we evaluate the ensemble average of the optical response over many stochastic realizations of atomic positions.

\subsubsection{Truncation of hierarchy and ``cooperative Lamb shift''}

We may obtain a MFT solution to the transmitted light by ignoring recurrent scattering processes and any light-induced correlations. In the main text this was first done for $N$ atoms  by calculating the product of the $N$ independent single-atom transmission results. Alternatively, we may obtain effective medium MFT solutions directly from the hierarchy of equations for the correlation functions by means of an appropriate truncation of the hierarchy. The
lowest-order recurrent scattering process where an atom pair is involved in the repeated photon exchanges appear in the equation for ${P}_{2}(x;x')$. We may ignore these, and any correlation effects between a ground-state atom $x$ and a polarization at $x'$, by factorizing the two-atom correlation function,
\beq
P_{2}(x_1;x_{2})= \rho(x_1) P_1(x_2)\,.
\label{decorre}
\eeq
When Eq.~\eqref{decorre} is substituted in the steady-state solution for $P_1$ [Eq.~(4) in the main text], or to Eq.~\eqref{p1app}, we obtain a closed equation for $P_1$,
\begin{equation}
P_1(x) = \alpha\rho \tilde{D}^+_F(x)+ \eta_\delta \rho(x) \int
dx' e^{ik|x-x'|}  P_1(x')\,.
\label{mf}
\end{equation}
The decorrelation approximation \eqref{decorre} is an effective medium theory where each atom interacts with the average behavior of its neighboring atoms and the position correlations due to the dependence of the dipole-dipole interactions on the relative interatomic separation $x_1-x_2$ is ignored. The
optical response that deviates from the solution of \EQREF{mf} therefore indicates a nonvanishing
value for light-induced two-atom correlations
\beq
\delta P_{2}(x_1;x_{2})\equiv P_{2}(x_1;x_{2})- \rho(x_1) P_1(x_2)\,.
\eeq

In a 1D waveguide we can solve (\ref{mf}) for the atoms filling the region $x\in[0,L]$ by substituting $P_1(x)=P_+ \exp(ik' x)+P_-  \exp(-ik' x)$, for $0\leq x\leq L$. Here $k'$ takes complex values [${\rm Im}(k')>0$], representing a damped plane wave propagation.
The comparison between \EQREF{mf} and the expectation value of \EQREF{eq:MonoD1d} yields $\alpha \rho \eo \<\tilde E^+(x)\> = P_1(x)$. By means of the refractive index $n=k'/k$, we can then express the electric susceptibility as
\beq
\chi=n^2-1=\alpha\rho\,.
\label{chi}
\eeq
Note that, unlike in the corresponding 3D system, the Lorentz-Lorenz local field correction is absent here.

In the low atom-density limit in a 3D system, the  factorization analogous to \EQREF{decorre} leads to the familiar expression for the ``cooperative Lamb shift'' (CLS) that was originally calculated by
Friedberg, Hartmann and Manassah for atoms in a 3D slab geometry~\cite{Friedberg1973}, and was recently experimentally verified in a hot atom vapor~\cite{Keaveney2012}. Here we calculate CLS and the corresponding resonance width for the 1D waveguide by solving \EQREF{mf}, subject to \EQREF{chi}. The low atom density expansion of MFT yields
\beq
\Delta_{\rm CLS} ={\gamma_w\rho\over 2k}\left( 1-\frac{\sin 2Lk}{2Lk}\right) +{\cal O}[\gamma_w^2\rho^2/(k^2\gamma_t)]\,.
\label{COLLAMBb}
\eeq
The oscillatory behavior results from the etalon interference effects of the sample thickness. Figure~3 in the main text shows example cases for the comparison between the CLS and the resonance shift of the full numerical solution. With carefully chosen parameter values the result is close to the CLS shift \eqref{COLLAMBb}. However, when, for instance, the optical thickness of the sample is increased, the oscillatory behavior is absent in the full numerical solution and the CLS result qualitatively fails.

The resonance HWHM width $\Gamma_{\rm FHM}$ can also be derived from the effective continuous medium MFT model to the same order in the density expansion as CLS. We find
\beq
\Gamma_{\rm FHM} =\gamma_t\sqrt{1+{\gamma_w\rho\over \gamma_t k} \( {1+ 2L^2 k^2 -\cos(2Lk)\over 2Lk} \)}\,. \label{appwidth}
\eeq
For thin atomic ensembles we may also expand the square root,
\beq
\Gamma_{\rm FHM} =\gamma_t+{\gamma_w\rho\over k} \( {1+ 2h^2 k^2 -\cos(2hk)\over 4hk} \)\,. \label{appwidth2}
\eeq

\subsection{Transfer matrix solutions}

\subsubsection{Basic relations}

In this section we provide a description for the light-induced correlations between the atoms in light propagation and how they depend on the density of the atoms and their thermal distribution.
The analysis utilizes transfer matrix theory where we adapt theoretical methods of localization analysis in transport phenomena~\cite{anderson1d} that were originally developed for describing electric conductivity in 1D wires.

For fixed atomic positions the atomic excitations can be solved in the low light-intensity limit from \EQREF{eq:classicaledb}. The total electric field amplitude is then obtained by substituting the solution into \EQREF{eq:pointlight}.
In order to analyze the transmission of light by one atom we rewrite Eqs.~\eqref{eq:pointlight} and~\eqref{eq:classicaledb} in the vicinity of the atom $l=1$ as
\begin{align}
\eo \tilde E^+(x) &= \eo \tilde E_{\rm ext}^+(x)+ {ik\over2}  e^{ik|x-x_1|} \mathfrak{ P}^{(1)},
\label{eq:tr1}\\
\mathfrak{ P}^{(1)} &= \alpha \eo \tilde E_{\rm ext}^+(x_1)\,,\label{eq:tr2}
\end{align}
where $\tilde E_{\rm ext}^+(x_1)$ represents the incident field plus the scattered fields by all the other atoms in the ensemble. We separate the external fields that propagate from the negative $x$ direction ($x_-<x_1$) to the atom and those that propagate from the positive direction ($x_+>x_1$) to the atom
\beq
\tilde E_{\rm ext}^+(x_1)= \tilde E_{{\rm ext},-}^+(x_-) e^{ik(x_1-x_-)}+\tilde E_{{\rm ext},+}^+(x_+) e^{-ik(x_1-x_+)}
\eeq
Similarly, we separate the total field $\tilde E^+(x) $ into the field propagating in the positive $x$ direction \emph{away} from the atom, $\tilde E^+_+(x) $, and the field propagating in the negative $x$ direction away from the atom, $\tilde E^+_-(x) $. Substituting these into Eqs.~\eqref{eq:tr1} and~\eqref{eq:tr2} and separating the components of the two propagation directions, we find
\begin{align}
\tilde E^+_+ &= \tilde E_{{\rm ext},-}^+ +\eta_\delta (\tilde E_{{\rm ext},+}^+ + \tilde E_{{\rm ext},-}^+),\\
\tilde E^+_- &= \tilde E_{{\rm ext},+}^+ +\eta_\delta (\tilde E_{{\rm ext},+}^+ + \tilde E_{{\rm ext},-}^+)\,.
\end{align}
We solve these equations for the field amplitudes for the region $x>x_1$ in terms of the amplitudes for $x<x_1$. The fields at $x_1+$ and $x_1-$ are related by
\begin{align}
&\left[
\begin{array}{c}
\tilde E^+_+ \\
\tilde E_{{\rm ext},+}^+
\end{array}
\right]
=
{\cal T}
\left[
\begin{array}{c}
\tilde E_{{\rm ext},-}^+\\
\tilde E^+_-
\end{array}
\right],\\
&{\cal T} (\delta) =  \left[
\begin{array}{cc}
 \frac{(2 \eta_\delta +1)}{\eta_\delta +1} & \frac{ \eta_\delta }{\eta_\delta +1} \\
 -\frac{\eta_\delta }{(\eta_\delta +1)} & \frac{1}{(\eta_\delta +1)} \\
\end{array}
\right],
\end{align}
where ${\cal T}$ is the transfer matrix for this problem.
For a homogeneously broadened system we may consider all the detunings $\delta_j$ of the light from the atomic resonance to be equal. For an inhomogeneously broadened thermal gas these fluctuate according to the Doppler broadening that results from the thermal Maxwell-Boltzmann distributed atomic velocities.

For a single atom the transmission and reflection amplitudes,  $t^{(1)}$ and  $r^{(1)}$ respectively, follow directly from the transfer matrix. On the right side of the atom there is only the transmitted wave, call its amplitude $\tilde E^+_t$, whereas on the left we have the incoming and reflected waves, $[\tilde E^+_i,\tilde E^+_r]^T$. These satisfy
\beq
\left[
\begin{array}{c}
\tilde E^+_{t} \\
0
\end{array}
\right]
={\cal T} \left[
\begin{array}{c}
\tilde E^+_{i} \\
\tilde E^+_{r}
\end{array}
\right]\,.
\label{eq:1atomtrans}
\eeq
We obtain
\begin{align}
t^{(1)} (\delta) &={\tilde E^+_{t}\over \tilde E^+_{i}}= {(\gamma_w-\gamma_t)+i\delta\over i\delta-\gamma_t},\\
r^{(1)}(\delta) & ={\tilde E^+_{r}\over \tilde E^+_{i}} =\eta_\delta ={\gamma_w\over i\delta-\gamma_t}\,,
\label{eq:singletrans_amp}
\end{align}
where we write
\beq
\eta_\delta = \sqrt{R^{(1)}} \zeta, \quad \zeta=e^{i\varphi}, \quad \varphi=\arctan(\delta/\gamma_t)\,.
\label{reflectionstuff}
\eeq
Here $\varphi$ denotes the phase factor associated with the reflection.
The transmission and reflection can be described in terms of the single-atom power transmission and reflection coefficients $T^{(1)} =|t^{(1)} |^2$ and $R^{(1)}= |r^{(1)} |^2$,
\beq
T^{(1)} (\delta)={(\gamma_t-\gamma_w)^2+\delta^2\over \gamma_t^2+\delta^2},\quad R^{(1)}(\delta) ={\gamma_w^2\over \gamma_t^2+\delta^2}\,.
\label{eq:singletrans_b}
\eeq

\subsubsection{Stochastic atomic positions and detunings}

In order to analyze the suppression of light-induced correlations and the emergence of the MFT response, we first consider the case of two atoms at $x_1$ and $x_2$. Here we also need to consider the propagation phases of the light from $x'$ to $x$, which are governed by the matrix
\beq
\Phi(x,x')  = \left[\begin{array}{cc}
e^{ik(x-x')} & 0\\
0&e^{-ik(x-x')}
\end{array}\right]
\eeq
for both the right- and left-propagating waves.

When considering amplitude transmission, ordinarily one is comparing the properties of the light between some fixed points before and after the sample, call them $x_0$ and $x_3$. We are then back to the same problem as in the case of a single atom, except that the composite transfer matrix has to include the two atoms and the propagation phases:
\beq
{\cal T}_{03}=\Phi(x_3,x_2) {\cal T}(\delta_2)\Phi(x_2,x_1){\cal T}(\delta_1)\Phi(x_1,x_0).
\eeq
We obtain the two-atom transmission amplitude
\beq
t_{12}^{(2)} = { t_2^{(1)} t_1^{(1)} \over 1 - \sqrt{R_1^{(1)} R_2^{(1)}} \zeta_1\zeta_2 \xi_{12}} e^{ik(x_3-x_0)}\,.
\label{t12}
\eeq
Here $\xi_{12} = \exp(2 i k x_{12})$ (where $x_{12}=x_2-x_1$) is a propagation phase associated with a back-and-forth trip of the light between atoms 1 and 2, and $\zeta_1$ and $\zeta_2$ are phase factors upon reflection from each atom. The overall phase factor for propagation from $x_0$ to $x_3$ is trivial, and is henceforth omitted.

We can write Eq.~\eqref{t12} as a geometric series expansion
\begin{align}
t_{12}^{(2)} &=  t_2^{(1)} t_1^{(1)}  + t_2^{(1)} \sqrt{R_1^{(1)} R_2^{(1)}}  e^{i\phi} t_1^{(1)} \nonumber\\& +   t_2^{(1)} \sqrt{R_1^{(1)} R_2^{(1)}}  e^{i\phi}  \sqrt{R_1^{(1)} R_2^{(1)}}  e^{i\phi} t_1^{(1)} + \ldots \,,
\label{t12exp}
\end{align}
with
\beq
\phi=\varphi_1+\varphi_2+2 k x_{12}\,.
\eeq
The interpretation of Eq.~\eqref{t12exp} is straightforward: the sum is over all the repeated photon exchanges between the two atoms, and each term $(R_1^{(1)} R_2^{(1)})^{1/2}$ represents one recurrent scattering event for the photon.

Next we calculate the ensemble averages for the transmitted light for the composite two-atom system. For the transmission amplitude we have
\beq
\<  t_{12}^{(2)} (x_{12},\delta_1,\delta_2) \>_{x_{12},\delta_1,\delta_2} =  \big\<{ t_2^{(1)} t_1^{(1)} \over 1 - \sqrt{R_1^{(1)} R_2^{(1)}} \zeta_1\zeta_2 \xi_{12}} \big\>\,.
\label{eq:ave1}
\eeq
The subscript indicates the averaging over the ensemble of interatomic separations $x_{12}$, and the detunings of the atoms 1 and 2, $\delta_1$ and $\delta_2$, respectively. The detuning $\delta_j$ appears in the single-atom transmission and reflection amplitudes  $t_j^{(1)}$ and $[R_j^{(1)}]^{1/2} $ (in fact $R^{(1)}_j\rightarrow 0$ with $\delta_j\rightarrow \infty$), as well as in the phase $\zeta_j$.
Similarly, we can calculate the expectation value of the optical thickness for the composite two-atom system
\beq
\<D_{12}^{(2)} \>=- \<\ln T_{12}^{(2)}\> = -\big\< \ln \big({  T_2^{(1)} T_1^{(1)}\over \big|1 - \sqrt{R_1^{(1)} R_2^{(1)}} e^{i\phi}\big|^2} \big)\big\>\,,
\label{eq:ave2}
\eeq
where
$T_{12}^{(2)}= | t_{12}^{(2)} |^2$  denotes the transmission coefficient for the intensity of the two-atom system.

If the atomic positions are random, the propagation factor $ \xi_{12}$ fluctuates according to the relative positions between the atoms 1 and 2. For a homogeneously-broadened gas we set all the detunings equal $\delta_1=\delta_2$. Consequently, the amplitudes $t_1^{(1)}=t_2^{(1)}$ and $[R_1^{(1)}]^{1/2} =[R_2^{(1)}]^{1/2} $, and the phase factors $\varphi_1=\varphi_2$ are constant. For an inhomogeneously broadened thermal gas the detunings are Doppler broadened as a result of the thermal Maxwell-Boltzmann distribution of the atomic velocities.
We then take $\delta_1$ and $\delta_2$ to be independent Gaussian distributed random variables, where the Gaussian distribution $f(\delta)$ is given by
\beq
f(\delta) = {1\over \sqrt{2\pi} \Delta\omega} e^{-{\delta^2 / (2 \Delta\omega) }}\,,
\eeq
with $\Delta\omega\equiv k \sqrt{k_B T/m}$ determined in terms of the wavenumber of light $k$, the temperature of the atoms $T$, the mass of the atoms $m$, and the Boltzmann constant $k_B$. The ensemble average of the transmitted light amplitude in a composite two-atom system then reads
\begin{align}
\<  t_{12}^{(2)} &(x_{12},\delta_1,\delta_2) \>_{\delta_1,\delta_2} \nonumber\\
&=  \int_{-\infty}^{\infty} d\delta_1 d \delta_2 {  f(\delta_1) f(\delta_2)  t_2^{(1)}(\delta_2) t_1^{(1)}(\delta_1)\over 1 - \sqrt{R_1^{(1)}(\delta_1) R_2^{(1)}(\delta_2)} e^{i\phi(\delta_1,\delta_2)}}
\label{eq:avether}
\end{align}
Provided that the width of the Gaussian distribution satisfies
\beq
k\sqrt{k_BT/m}\gg \gamma_t\,,
\eeq
the contribution of the second term in the denominator in \EQREF{eq:ave1} vanishes. We then find that the transmission amplitudes of the atoms 1 and 2 decouple
\beq
\< t_{12}^{(2)}\>_{\delta_1,\delta_2} =  \< t_1^{(1)}\>_{\delta_1}  \< t_2^{(1)}\>_{\delta_2}  \,.
\label{eq:corrcan}
\eeq
The optical thickness of the two-atom system similarly decouples to the optical thicknesses of the individual atoms,
\beq
\< D_{12}^{(2)}\>_{\delta_1,\delta_2}  =    \< D_2^{(1)}\>_{\delta_2}  + \< D_1^{(1)}\>_{\delta_1} \,.
\label{eq:corrcan2}
\eeq
The emergence of MFT is a general consequence of inhomogeneous broadening, and is not necessarily due to thermal atomic motion. For quantum dots or circuit resonators
it can result, e.g., from fabrication imperfections~\cite{JenkinsRuostekoskiPRB2012b}.

The decoupling \eqref{eq:corrcan} [or \eqref{eq:corrcan2}] can be achieved even in a homogenously broadened sample when $\delta_j$ are constant, if the propagation phases $2k x_{12}$ sufficiently fluctuate [$\Delta(x_{12})\gg \pi/k$, where $(\Delta x)^2=\<x^2\> - \<x\>^2$]. In a homogeneously broadened two-atom system, $\delta_1=\delta_2=\delta$, we have for completely randomly distributed $2k x_{12}$
\begin{align}
\<  t_{12}^{(2)} (x_{12},\delta,\delta) \>_{x_{12}} &= {1\over 2\pi} \int_0^{2\pi} d\phi {  t_2^{(1)} t_1^{(1)}\over 1 - \sqrt{R_1^{(1)} R_2^{(1)}} e^{i\phi}} \nonumber\\
&= t_2^{(1)}  t_1^{(1)}  \,.
\label{eq:average1}
\end{align}
If the phases $2k x_{12}$ are approximately evenly distributed over the entire interval $[0,2\pi)$, the second term in the denominator in \EQREF{eq:average1} is of no consequence. Analogously, we find
\beq
\<  D_{12}^{(2)} (x_{12},\delta,\delta) \>_{x_{12}} = \< D_1^{(1)}\>_{x_{12}}  +  \< D_2^{(1)}\>_{x_{12}}.
\label{another_average}
\eeq

The two-atom results \eqref{eq:corrcan}--\eqref{another_average} may be recursively generalized to $N$ atoms.
Whichever is the cause of strong averaging that eliminates recurrent scattering, we have
\begin{align}
\<t^{(N)}_{1,2,\ldots,N}\> & =\< t_1^{(1)} \>\< t_2^{(1)} \>\ldots \< t_N^{(1)} \>,  \label{eq:simpleTMa}\\
\<D^{(N)}_{1,2,\ldots,N} \> &= \<D^{(1)}_{1} \> +  \<D^{(1)}_{2} \>+\ldots + \<D^{(1)}_{N} \> \,,
\label{eq:simpleTM}
\end{align}
where $D^{(1)}=-\ln T^{(1)}$ and $T^{(1)}$ is given by \EQREF{eq:singletrans_b}.

Owing to the total reflection of light by a resonant atom whenever $\gamma_w=\gamma_t$, the optical thickness $D = -\ln T$ diverges on resonance. The divergent optical thickness can be cured if we consider a Poisson-distributed number of atoms in a waveguide, and average over the atom number. We find for the MFT result for the optical thickness with the average atom number $\bar N$
\beq
D(\bar N) = -\ln T^{(N)}=\frac{(2\gamma_t -\gamma_w )\bar N \gamma_w }{\gamma_t^2+\delta^2}\,.
\label{DPOISS}
\eeq
The expression is finite, since there is a nonvanishing probability of having zero atoms in the waveguide, so that the average transmission is never zero.

The simplicity of the full solutions \eqref{eq:simpleTMa} and \eqref{eq:simpleTM} is remarkable. Even though the calculation incorporates all the recurrent scattering processes where the photons are repeatedly exchanged between the same atoms, in the final result they have precisely canceled out. The exact cooperative response of the atomic ensemble coincides with the approximate MFT analysis in which each atom in a row simply passes on the same fraction of light that it would if there were no other atoms present. The MFT solution neglects all the light-induced correlations between the atoms. These correlations are suppressed by the fluctuating detunings of the light from the atomic resonances or phase factors associated with each atom in the scattering processes. In particular, the exact solution and the MFT result coincide whenever these fluctuations are sufficiently strong. We find that the first condition is true whenever the Doppler broadening considerably exceeds the resonance linewidth of the atoms. The fluctuations in the propagation phase factors due to the relative atomic positions, on the other hand, are sufficiently strong for the MFT to be valid if the atom density is low enough.

In order to illustrate the cancelation of the effect of recurrent scattering, we first calculate the effect of thermal broadening on the two-atom correlation function.
In Fig.~\ref{fig:thermal}(a) we show the relative deviations $r$ of the exact two-atom result from the two-atom MFT result as a function of temperature. Here we have defined
\beq
r=  \left| { \< t_{12}^{(2)}\>_{\delta_1,\delta_2} -    \< t_2^{(1)}\>_{\delta_1}\< t_1^{(1)}\>_{\delta_2} \over     \< t_2^{(1)}\>_{\delta_1}\< t_1^{(1)}\>_{\delta_2} } \right| \,.
\label{eq:relative}
\eeq
For simplicity, we consider constant relative atomic positions by setting $2kx_{12}=0$ in Fig.~\ref{fig:thermal}.
The MFT result  corresponds to the case where the light-induced correlations between the two atoms are ignored.

\begin{figure}
\includegraphics[width=0.7\columnwidth]{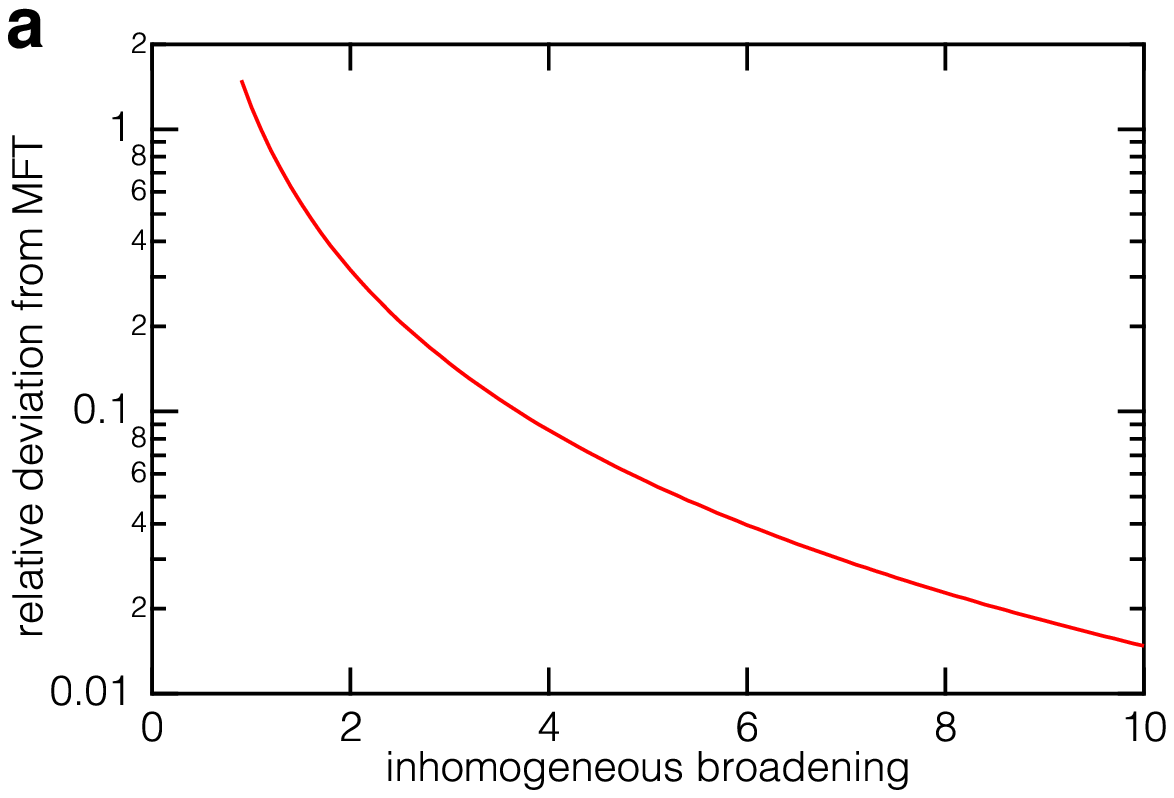}
\includegraphics[width=0.7\columnwidth]{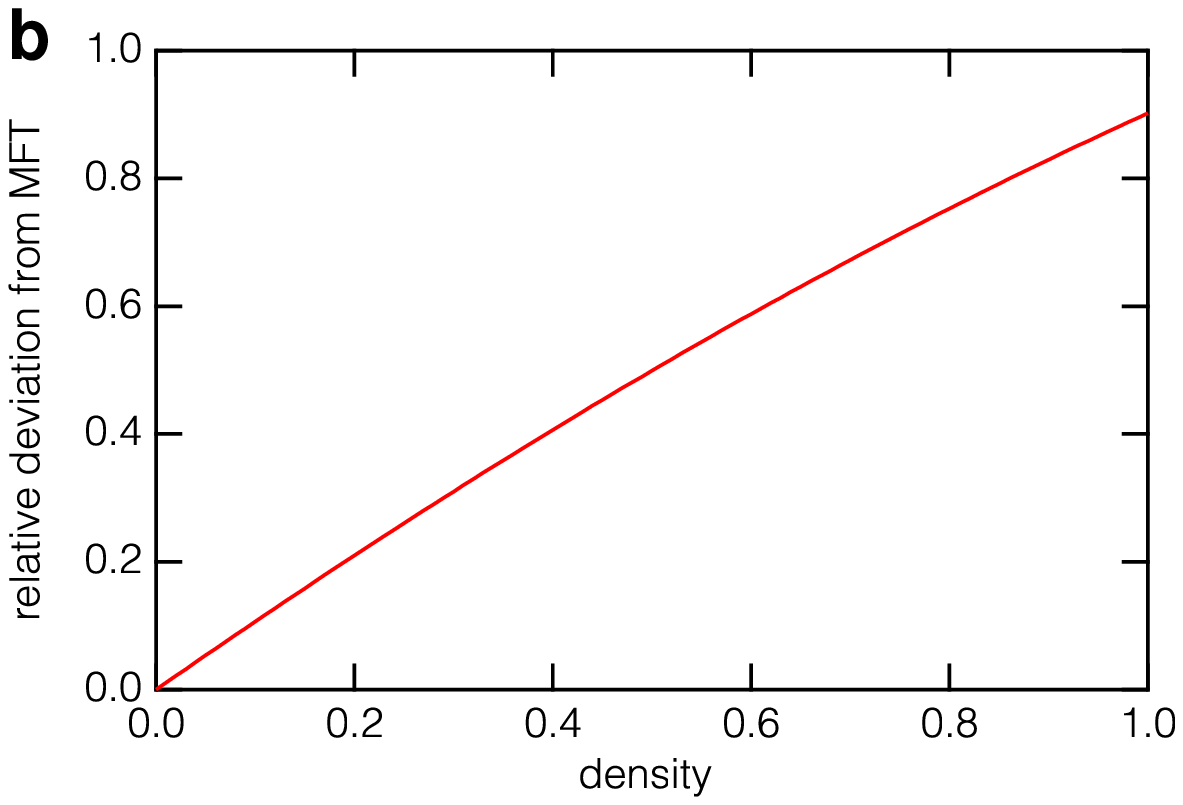}
\caption{Relative deviation of exact two-atom transmission amplitude from the corresponding MFT result. (a) The deviation $r$ [\EQREF{eq:relative}] as a function of  the temperature, which is expressed as the corresponding inhomogeneous broadening in units of the linewidth of the atomic transmission $k\sqrt{k_BT/m}/\gamma_t$. This panel is for the extreme case when the fixed back-and-forth propagation phase between the two atoms is a multiple of $2\pi$. (b) The deviation $r$ for a homogeneously broadened two-atom system with $\delta/\gamma_t=0.1$ as a function of the dimensionless density $\rho/k$.  There are no losses from the fiber in either case, so we have $\gamma_w=\gamma_t$.}
\label{fig:thermal}
\end{figure}

Next we analyze the effect of the atom density on the cancelation of the effect of recurrent scattering in a homogeneously-broadened system.
We may estimate the distribution of the relative positions of adjacent atoms in a classical, initially uncorrelated atomic ensemble. Let us assume that there is a number of atoms $n\gg 1$ distributed over the interval $\Delta x$ for $x>x_1$ where $x_1$ is the position of the atom 1. The probability that there are no atoms on the interval $[x_1,x_1+\delta x]$ is then $(1-\delta x/\Delta x)^n\rightarrow e^{-\rho \delta x}$, when $n\rightarrow\infty$, where the density $\rho=n/\Delta x$.
In the limit of large $n$, we may replace $n$ and $\Delta x$ by $N$ and $L$, respectively.
The $1/e$ width of the probability distribution of no atoms adjacent to the atom 1 for $x>x_1$ is
$\Delta(x-x_1)\simeq1/\rho$.
The condition for the sufficient fluctuation of the propagation phases $2k x_{12}$ for Eqs.~\eqref{eq:simpleTMa} and~\eqref{eq:simpleTM} to be valid is then approximately given by
\beq
\rho/k\ll 1/\pi\,.
\eeq
We can then analytically calculate the expectation value of the two-atom transition amplitude for a homogeneously broadened atom pair by using the distribution
\beq
g(\phi)= \frac{\rho}{2k}\, e^{-\rho \phi/2k},\quad 0\leq \phi <\infty\,,
\eeq
for the back-and-forth propagation phase $\phi=2 k(x-x_1)$.
But, if we wish to calculate the average of a $2\pi$ periodic function $F(\phi)$, by virtue of the specific form of $g(\phi)$ we have
\begin{align}
\<F(\phi)\> &= \int_0^\infty d\phi\,g(\phi)F(\phi)\nonumber\\ &=
\sum_{m=0}^\infty \int_{2\pi m}^{2\pi(m+1)} d\phi\, {g}(\phi)F(\phi) \nonumber\\ &=\sum_{m=0}^\infty e^{-m\pi\rho/k} \int_0^{2\pi} d\phi\, {g}(\phi)F(\phi)\nonumber\\
&=\int_0^{2\pi}d\phi\,\tilde{g}(\phi)F(\phi)
\end{align}
where $\tilde{g}$ is simply the probability density $g$ normalized over the interval $[0,2\pi)$,
\beq
\tilde{g}(\phi)= {\rho/k\over 2(1- e^{\pi\rho/k} )}\, e^{-\rho \phi/2k},\quad 0\le\phi < 2\pi\,.
\eeq
We therefore find that
\begin{align}
\< & t_{12}^{(2)} (x_{12},\delta,\delta) \>_{x_{12}}  = \int_0^{2\pi} d\phi\, {\tilde{g}(\phi)   t_2^{(1)} t_1^{(1)}\over 1 - \sqrt{R_1^{(1)} R_2^{(1)}} e^{i\phi}}\nonumber\\
&=  { [\delta -i (\gamma_w-\gamma_t)]^2 \over  (\delta+i\gamma_t)^2} \, \mbox{}_2 F_1 (1, {i\rho\over 2k}; 1+  {i\rho\over 2k}; - {\gamma_w^2\over (\delta+i\gamma_t)^2})
\,,
\label{eq:analytic}
\end{align}
where $\mbox{}_2 F_1$ denotes the hypergeometric function.
In Fig.~\ref{fig:thermal}(b) we show the relative deviations $r$ [defined analogously to \EQREF{eq:relative}, but with the averaging now with respect to $x_{12}$] of the exact two-atom result from the two-atom MFT result as a function of the density of the atoms. The exact two-atom result \eqref{eq:analytic} qualitatively differs from the MFT value when the density approaches to $\rho\simeq k$.

The uneven distribution of the propagation phases at high atom densities can also be illustrated by simple numerical examples.
For instance for a density $\rho/k\simeq 0.06$ there is a less than 9\% chance for the propagation phases to satisfy $2k x_{12}<\pi$. However, for the density $\rho/k\simeq 2$ this probability is increased to 96\%, indicating that the phases are then expected to accumulate substantially more in the interval  $[0,\pi)$ than in $[\pi,2\pi)$, leading to a notably uneven phase distribution. In such a case the averaging resulting in the MFT response of Eqs.~\eqref{eq:simpleTM} and~\eqref{DPOISS} is no longer valid, and the light transmission is affected by light-induced correlations between the atoms.

\subsubsection{Waveguide with a strong loss rate}

We can also consider the limit $\gamma_w/\gamma_t\ll 1$. The two-atom transmission is then to leading order in $\gamma_w/\gamma_t$
\beq
T_{12}^{(2)} \simeq  T_1^{(1)} T_2^{(1)}\left(1 - {\cal O} (\gamma_w^2/\gamma_t^2)\right)\,,
\eeq
and to the lowest order we again recover the MFT response. Figure~2 in the main text illustrates how the exact numerical solution approaches the MFT result as the loss rate of the photons from the waveguide is increased.

\end{document}